\documentclass[preprint,12pt]{aastex}
\def\ltw{\>\hbox{\lower.25em\hbox{$\buildrel <\over\sim$}}\>}
\def\gtw{\>\hbox{\lower.25em\hbox{$\buildrel >\over\sim$}}\>}
\def\be{\begin{equation}}
\def\ee{\end{equation}}
\shorttitle{M87: A Different Picture}
\shortauthors{Owen, Eilek, Kassim}

\begin{document}
\title{\bf M87 at 90cm: A Different Picture}

\author{Frazer N. Owen}
\affil{National Radio Astronomy Observatory\altaffilmark{1}} 
\affil{P. O. Box O,Socorro, NM  87801}
\email{fowen@aoc.nrao.edu}

\author{Jean A. Eilek}
\affil{Physics Department, New Mexico Tech}
\affil{Socorro, NM 87801}
\email{jeilek@aoc.nrao.edu}

\and

\author{Namir E. Kassim}
\affil{Naval Research Lab}
\affil{Washington, DC 20375}
\email{kassim@rsd.nrl.navy.mil}

\altaffiltext{1}{The National Radio
Astronomy Observatory is operated by  Associated Universities, Inc., under
a cooperative agreement with the National Science Foundation.}
\begin{abstract} 
	We report new radio imaging of the large scale radio structure
of M87 with the VLA at 90 cm. These new images show the complex structure
of the radio emission more clearly than previous attempts, some of
which date back to 
the 1940's. The images suggest that the outward flow from the M87 nucleus
extends well beyond the 2 kpc jet. Two ``bubbles'' of synchrotron 
emission appear
to be  inflated by this flow. A simple model of the emission, combined
with our knowledge of the inner jet, suggests that the energy input into
to this region from the M87 nucleus exceeds the energy being radiated
away as X-rays. This argues that the region within 40
kpc of the center of M87 is currently dominated by energy input from the
M87 nucleus. The gas in the region is expanding, not flowing inward as
is envisioned in the cooling flow model. 
\end{abstract}


\clearpage

\section{Introduction}

	M87 is one of the most famous radio galaxies, mainly due to
studies of its  non-thermal jet and activity in its nucleus. 
However, most of the radio emission responsible for its discovery in
the 1940's \citep{b49} comes from a much larger scale with a projected
end-to-end length of about 80 kpc \citep{m52,bm54}. The large scale
structure has been studied more recently \citep{f87,k93,b95,ro96}.
Nevertheles, the connection of the larger radio structure 
to the jet and the inner lobes has remained obscure.  

M87 sits at the center of the X-ray luminous atmosphere of the Virgo cluster
\citep{flg}.  The X-ray atmosphere has a simple,
apparently undisturbed morphology with a central luminosity peak.
This atmosphere has been interpreted as a cooling flow
\citep{s84}, but the situation may be more complex
\citep{bi99}.  In addition, the inner region of the X-ray atmosphere has
an unusual structure \citep{b95, h99}.  Once again, the  
connection of the radio source to the X-ray atmosphere has
remained obscure.  

	In this paper we report new 90 cm observations with the VLA of
the large scale radio emission with much higher dynamic range and
resolution than has been achieved previously.   Based on these new
data, we propose that the radio halo is alive, rather than being a relic
of some previous activity,  and it is interacting
strongly with the inner regions of the cluster's X-ray atmosphere.

\section{Observations \& Results}

	The observations were made over a period of
many years with the VLA in spectral line mode. The earlier data comes
from dual frequency observations at 74 and 321 MHz in the A, B and
C configurations. Images from these data were encouraging, so the
bulk of the data in the current image were collected in the 1998
A and B arrays at 321 and 327 MHz, also in the spectral line mode.
The use of narrow channels of 0.1 MHz or less reduces the systematic
instrumental errors, which often limit such observations, and the
radial smearing of the confusing sources, which makes them easier to
remove and scatters less real power into the noise. It also facilitates 
editing of intermittent, mainly external, narrow-band RFI.

	The observations used 3C286 for initial amplitude and phase 
calibration and to set the standard VLA flux density scale which
derives from \citet{b77}. The entire field was imaged using the AIPS
program IMAGR in 3D mode and selfcalibrated using CALIB. The confusing
sources were then subtracted from the {\it uv} data and the remaining
imaging and calibration was performed on just the central  field
containing M87. Further editing, imaging and self-calibration were
then   carried out. The final image was produced using the maximum
entropy program, VTESS. In figure \ref{m87color}, we show the resulting
image as a false-color representation of the full image. In figure 
\ref{m87pc}, we show a grey-scale blowup of the
central region showing the complex structure of brighter features more
clearly. In figure \ref{minpfig} we show a greyscale of the full source
as well as the locations of minimum pressure measurements discussed later
in the paper. 

Several features of this image are striking\footnote{see also
\citet{kl99} for an overall description of the Virgo A halo.}. The very inner
region (which appears 
as the red/orange patch in the center of figure \ref{m87color}) contains the
well-known jet, which points to the northwest.   This inner region (or
inner ``lobes'')
extends about 2 kpc  from the core.   Two collimated flows
emerge from this region, one directed almost exactly eastward, and the
other directed slightly north of west.  The initial direction of the
westward flow is roughly aligned with the direction of the inner jet,
although the flow quickly bends and twists once it leaves the inner
region.  The flows appear to consist of a mass of bright, curved
structures, which we call filaments.  The eastern flow ends in a
well-defined pair of edge brightened circular lobes.  This ear-shaped
structure is reminiscent of a subsonic vortex ring.  The western flow
develops a gradual but well-defined southward twist, starting only a
few kpc beyond the inner lobes.    Finally, both flows are immersed in
a larger structure which might be described as two overlapping
``bubbles'',  each extending about 40 kpc from the nucleus.  The
coherent flows appear to be continuous, from the point at which they
emerge from the inner lobes, to the outer edge of the radio halo.
After reaching the halo, the flows gradually disperse -- the westward
flow particuarly -- and appear to be filling the entire halo with
radio-loud, filamented plasma.  

	As can be seen in figure \ref{m87color}, the bubbles end abruptly,
with well-defined outer boundaries which are slighly limb brightened
around perhaps half of the periphery.  The brightest filaments stand
out clearly from the general emission, suggesting higher minimum
pressures than the average over the source.  


	The results of minimum pressure analysis for the magnetic fields
and relativistic particles in the filaments are
given in Table \ref{minp} as calculated at  the positions indicated in
Figure \ref{minpfig}. We have assumed a frequency range of 10 MHz
to 10 GHz; a spectral index, $\alpha=1.0$ ($S\propto \nu^{-\alpha}$),
a proton-to-electron energy ratio $k=1$, and that the factor
$\zeta \phi = 1$, where $\zeta$ 
is the ratio of the true magnetic pressure to $B^2 / 8 \pi$, and $\phi$ is the
fraction of the volume occupied by relativistic electrons that is also
occupied by magnetic field. For the geometry of each feature we have
assumed a cylinder parallel to the line of sight
with radius equal to the VLA clean beam FWHM and a length
equal to the projected width on the sky ($W$).   We take a distance of
17 Mpc, so that 1 asec corresponds to 85 pc. 

	In table \ref{xrayp} we give estimated values of
the density, temperature and pressure from \citet{nb95} scaled to our assumed
distance of 17 Mpc. 
Comparison of tables \ref{minp} and \ref{xrayp} shows that all
the filaments have somewhat lower minimum pressures than the surrounding 
thermal gas. If the filling factor, $\zeta \phi$, is near one, the
more diffuse regions 
of the source have minimum pressures an order of magnitude below  
the thermal pressures. This could imply that the synchrotron emitting
regions are far from the minimum pressure case, that the filling
factor of the emitting regions is $\ll 1$, or that the thermal gas is
mixed with synchrotron emitting regions and provides much of the pressure
inside the filaments.

\section{Physical Picture:  the Radio Source}

	Virgo A does not fit neatly into the morphologies usually
assigned to radio galaxies.  It is clearly
not an FR type II, because it does not have outer hot spots.
  This has led people to describe it as an FR type
I.  While it may fit the strict definition of an FR I (being brightest
closest to the core), it has little in common with most FR I radio
galaxies.  Type I sources generally contain coherent jets which
slowly become wider and fainter the further out one looks from the
parent galaxy nucleus.  Virgo A, at least as we see it projected on
the sky, does not match this description.  

We suspect that Virgo A is more closely related to a class of 
amorphous, steep spectrum sources, e.g. \citep{b90}, 
which are found only in dense cores of galaxy clusters.  We are aware
of several comparable radio sources, e.g. in A133, A2052, 
A2626 \citep{r00}, and A2199 \citep{oe98};  all are attached to central
dominant galaxies in strong cooling cores. The fact that all such
sources are found in a special position relative to their parent
cluster (as well as their parent cluster being special in having a strong
cooling core) argues against them being simply normal, FR type I, tailed
sources seen in severe projection.  It seems more likely  that the
interaction of the jet with the dense, high-pressure external 
medium is more extreme than in most radio galaxies, and that the jet is
bent or disrupted as it tries to flow through the dense external medium.
The Virgo A radio halo is our best opportunity to study such an 
interaction.  	

\subsection{Nature of the Halo}

The M87 radio  halo, Virgo A, looks like  two bubbles which 
are supported by
 the energy  of poorly collimated outflows from the core.   Three
observations support this idea.  First, the radio halo
is the same size at all observed frequencies, from 74 and 330 MHz
\citep{k93}, to 10.6 GHz \citep{ro96}. 
Second, the outer boundary is relatively sharp and is slightly
limb brightened over about half of its circumference. Third, the
new images show what seem to be two directed outflows which connect the
inner jet and inner lobes to the full scale of the halo. 

From this evidence we can immediately
eliminate two possible models.  The halo is not due 
to single-particle diffusion through the cluster gas 
\citep[as in][]{an79, de80, d80}. Such a picture  predicts the size
would depend on 
frequency, due to synchrotron aging.  Nor is the measured size of the
halo limited by fading surface brightness in a simple outflow, as in
many Type I radio tails. This hypothesis 
would again predict frequency-dependent sizes, and also disagrees with
the sharp rise at the edge, described in \S 2. 

It is also apparent that the halo region is
inhomogeneous.  The halo plasma is highly filamented.
The radio-loud filaments can be regions with high field strength, high
relativistic particle density, or both.  We cannot uniquely
describe the interfilament regions.
 They could simply be low-field, or low-particle-density, 
regions.  Alternatively, they could be regions dominated by thermal
plasma from the cluster atmosphere.  Despite the fact that 
the outer edge of the radio halo is sharp and well defined, 
there is no strong evidence of a ``hole'' in the X-ray
image.\footnote{There is a weak feature apparent in the ringed X-ray
flux at $\sim 50$ kpc, suggestive of some interaction;  we have found
this in archived ROSAT data, and it can also just be seen in the
density profile from \citet{nb95}.}  This suggests that the X-ray loud 
plasma has managed at least partially to penetrate the halo region.

We draw an important conclusion from the image:  the halo is not
simply a relic of previous activity \citep[as in][]{tur75}. Rather, it is 
currently ``alive''.  It is being 
supplied with energy and relativistic plasma which come 
ultimately from the active nucleus and innermost radio jet of M87.
The well-directed inner jet has been disrupted on a scale of a few
kpc, but its mass and energy continue to flow out in ``plumes'' which
are less ordered, less well confined, than the flow interior to the
disruption. 

We emphasize that this system cannot be
static.  Continuing energy input from the radio jet must cause the
radio halo to expand outwards, into the cluster gas.  We propose the
following model of the system.  The radio halo is rather like a
bubble, with a 
well-defined outer edge which separates the internal, radio-loud 
plasma from the external, thermal cluster gas.  As the jet pours
energy into the halo, the pressure if the halo will rise, forcing it
to expand into the surrounding cluster gas.   The  observations show
that the halo is in
approximate pressure  balance with the surrounding cluster gas;  there
is no evidence from the radio for the halo being overpressure, and an 
underpressured region would quickly collapse.  Thus, we can model the
expansion can by assuming the halo pressure remains comparable to the
external pressure \citep[as in][]{es89}.  We will show below that
the expansion is slow and subsonic, so that the outer edge should not
involve any shock structures.  

A comment on the inner halo is in order.  The inner halo \citep{tur75,
hines89} is a small, bright region, approximately contiguous to the jet
and extending only $\sim 3$ kpc from the galactic core.  Previous
workers have assumed that the inner halo is the only active region
of the source.  There is, indeed, a sharp drop in surface brightness
between the inner and outer haloes, making imaging of the transition
very hard.  However, there is no observational evidence for a
well-defined outer edge to the inner halo, such as there is for the
outer edge of the outer halo. 

We thus differ with recent models of M87 \citep{r96,bb96}, which
assume the inner halo traps all  of the energy coming through the jet.
We can imagine 
two possibilities for the inner halo.\footnote{We defer quantitative
discussion of the inner-outer halo connection to a future paper.}  One
is that it is a quasi-steady part of the flow, resulting from some
choking of the flow which leads to a high-pressure, radio-bright inner
region.  If this is the case, the inner halo must be a porous
structure, with most of the jet power flowing through it to the outer
halo.  Another possibility is that the inner halo is a recent
transient, short lived compared to the outer halo (which we show below
is $\sim$ 100 Myr old), possibly due to a recent change in the
activity level of the core.  If this is the case it may have trapped a
significant part of the recent output from the jet, and be acting like
a small bubble within a larger bubble.  

Returning to the outer halo, and our picture of it as a large
expanding bubble, what can we say about the details? The magnetic
fields and relativistic  particles have minimum pressures somewhat
less than the thermal gas. However,  the uncertainities in such
analyses, especially the filling factor of the  synchrotron emitting
plasma, allow the possibility of a much larger relativistic
component. Thus, there are two possible models for what we are seeing,
depending on whether the region is dominated by thermal or
relativistic pressure. 

\subsubsection{Relativistically Dominated}

It is possible that the synchrotron emitting regions are far from the minimum
pressure case, with either the field or the relativistic particles strongly
dominating.  This can happen if the radio-loud region is 
inhomogeneous on scales smaller than the resolution of our image (that is,
if the radio-loud plasma has a small filling factor within our resolved
filaments), or if the relativistic protons carry much more energy than
do the relativistic electrons.   It can also happen 
if the region is simply dominated by pressure from
one of the components, either relativistic particles or magnetic field. 
If this is the case, then Virgo A is primarily a relativistic bubble.
The boundary apparent in the images  is between the primarily 
relativistic, radio-loud plasma and the 
thermal plasma further out in the cluster.   The ongoing energy input to
the plasma within this region will drive an expansion of the entire
bubble. 

If this is the case, thermal material inside the bubble  must come
from leakage through the boundary over the period of expansion.   We are
reminded of the semi-porous nature of the terrestrial magnetopause, which
allows partial penetration of the solar wind through local, intermittent
magnetic reconnection.  If this is the case, the mixing of thermal and
relativistic plasma internal to the halo is most likely macroscopic --
resulting in relativistic, radio-loud filaments with only low thermal
gas density, surrounded by interfilament regions dominated with thermal gas. 

\subsubsection{Thermally Dominated}

It is also possible that the synchrotron emitting regions are
 close to the minimum pressure case.  The results of the minimum pressure
analysis, taken at face value, are most consistent with a thermally
dominated model. In order to maintain pressure balance
with the X-ray atmosphere, the emitting regions must contain a significant 
amount of thermal gas.  
If this is so, the emitting regions represent only a small
fraction of the total energy inside Virgo A. In this case the jet
must deposit its material and energy directly in the thermal material.
 This picture requires more effective mixing,
on the microscopic scale, of the relativistic radio-loud plasma and the
ambient thermal plasma.   


If this is the
 case, the simple expanding bubble model can still describe the more
complex physical processes which are taking place. One possibility
is that the jet is depositing its matter, and energy, 
on scales $\sim$ tens of kpc from the core, in the form of turbulence and 
heat. Some of the turbulent energy may then be transformed into magnetic energy
through the turbulent dynamo process. The heated thermal gas will
expand to maintain balance with the unheated IGM outside this region.
Thus the bubble is formed.  The well-defined edge seen in the images is
then the interface between the heated, expanding, mixed thermal-relativistic
gas, and the undisturbed X-ray atmosphere ahead of the expansion surface.

\subsection{A Simple Dynamical Model}

Whichever picture best describes the M87 halo, the energy input
into the bubbles comes from the energy carried in the outflow from the
nucleus.  
We can use a simple model to describe the dynamics and estimate the
age of Virgo A.  
Following our analysis of A2199 \citep{oe98}, we model Virgo A as a
single spherical
bubble, with a steady energy input given by $P_j$, the total energy
flux from the nucleus and jet. We assume the bubble is in 
approximate pressure balance with the ambient medium, and  expands due
to its internal energy. This system can be analyzed through simple
energetics, without needing  detailed knowledge of the composition of
the plasma inside the bubble. 

\subsubsection{Jet Power}

We do need to estimate the jet power, however.  This is easier to do
for M87 than for more distant, less well-studied radio sources.  The
total flux of useable energy carried by matter in a relativistic jet
is  \citep[e.g., ][]{bb96}  
\be
P_j = \pi r_j^2 v_j \left[ \gamma_j (\gamma_j-1) \rho_j c^2 + 4
\gamma_j^2 p_j \right]
\ee
if $p_j, \rho_j, \gamma_j$ and $v_j$ describe the jet pressure,
mass density, Lorentz factor and velocity, respectively. We do not
know enough about the jet in M87 to estimate a good value for $P_j$,
but  we do know enough about  the jet to estimate its minimum
value, $P_{j, min}$.  We do this by considering only the internal
pressure in the jet plasma, which can be constrained by observations.
We thus ignore both the plasma inertia, $\rho_j c^2$, and any magnetic
field  exceeding that corresponding to the minimum-pressure
measurement.  (\citet{bb96} also argue from their specific instability
models that $\rho_j c^2 \ltw 4 p_j$.)  

The flow velocity in the inner jet, on scales $\ltw 2$ kpc,
 is very likely relativistic, with $\gamma_j \sim 3-5$ \citep{bzo95}.
 We can thus estimate $v_j \sim c$ and $\gamma_j \gtw 3$. The jet
radius can be measured directly from high-resolution images, such as
in \citep{ohc} (OHC). The jet pressure cannot be
found directly, but we can use the minimum pressure derived from
radio and optical observations.  \cite{bsh91} (BSH) derive minimum
 pressures by assuming the jet is uniformly 
 filled with radio-loud plasma.  Alternatively, OHC argue that
 the radio emission comes from thin helices on the surface of the jet;
 this small volume gives them minimum pressure values several times
 higher than given by BSH.  (Their model would require a
 comparable pressure throughout the jet, in order to maintain a stable
 structure.)   We can thus use the BSH value to
estimate the minimum likely $P_j$, noting that it may be higher.  
For knot D, we measure a radius of 22 pc;  BSH give
 $p_{min} \simeq 5.3 \times 10^{-9}$dyn/cm$^2$, and OHC give $p_{min}
 \simeq 1.2 \times  10^{-10}$dyn/cm$^2$.  For knot A,  we measure a
 radius of 62 pc;  BSH give 
 $p_{min} \simeq 3.5 \times 10^{-9}$dyn/cm$^2$, and OHC give $p_{min}
 \simeq 1.5  \times  10^{-10}$dyn/cm$^2$.   These result in a minimum
 jet power $P_{j,min} 
 \sim 7.8 \times 10^{43}$ erg/s at knot D, and $ \sim 4.7 \times
 10^{44}$ erg/s at knot A.  We conclude that a  conservative estimate
 of the jet power is $P_j \sim few \times  10^{44}$ erg/s,  and we use
 this to scale our modelling in what follows.  

Finally, we compare this to the total power radiated by M87.  The total
radio power of the source is $9.6 \times 10^{41}$ erg/s 
\citep[10 MHz - 150 GHz, scaled to 17 Mpc distance]{hr92}.  This is not the
total luminosity, however;  the jet is a strong optical and X-ray
source.  From BSH, we estimate the total jet power $\sim 2.8 \times
10^{42}$ erg/s; giving a total luminosity $\sim 3.7 \times 10^{42}$
erg/s for M87.  Interestingly, this is $\sim$ 1\% of our
conservatively estimated jet power, which in consistent with the
general picture of radio galaxies as  low-efficiency radiators.  

\subsubsection{Expansion and Age of the Outer Halo}

We picture the halo as a region of hot plasma, undergoing steady
energy input from the jet at rate $P_j$.  Let the halo plasma have
adiabatic index $\Gamma$, pressure $p$, volume $V$ and internal energy
$U_{int}$.  Radiative losses from the halo are $L_{rad}$. Energy
conservation tells us that the expansion is 
governed by \citep{es89}
\begin{equation}
{ d U_{int} \over dt} = P_j - p { d V \over dt} - L_{rad}
\end{equation}
If $L_{rad}$ is small compared to $P_j$ (as is true for M87), this
relation can easily be solved  for $V(t)$. 
To be specific, we assume the halo is a spherical bubble of radius
$R$, containing plasma with adiabatic index $\Gamma$, expanding 
due to its own internal energy.  In a slow expansion, during which the
pressure of the bubble does not greatly exceed that of the
surroundings, we can make the approximation that the inteior pressure
remains comparable to the exterior pressure.  If the exterior pressure
decays with radius, the governing equation becomes
\begin{equation}
4 \pi R^2 p_x(R)  { dR \over dt} = { \Gamma - 1 \over \Gamma} P_j
\end{equation}
The X-ray data, summarized in Table 2, can be represented as 
$p(r) \simeq 4.3 \times 10^{-11} ( r / r_o)^{-0.8}$ dyn cm$^{-2}$,
with $r_o = 12$ kpc. Using this in equation (2), we find 
\begin{equation}
{4 \pi \over 2.2}  { \Gamma \over \Gamma -1} p_o r_o^{0.8} R^{2.2} = P_j t
\end{equation}
We scale the jet power to $10^{44}$erg/s.   We take 
$R \sim 35$ kpc as the current size of the source.\footnote{The
projected semi-major axis of the halo is 37 kpc, and the semi-minor
axis is 22 kpc.  These are equivalent to a spherical volume of radius
30 kpc;  we choose $R = 35$ kpc to allow for modest projection
effects.}   

Details of the expansion are  sensitive to $\Gamma$, the internal
adiabatic index.   The two possibilities are  $\Gamma = 5/3$, which
describes a non-relativistic plasma (assuming most of the plasma in
the bubble is thermal), and $\Gamma = 4/3$
(appropriate if the bubble is dominated by relativistic particles  and
magnetic field).  The current age of the radio halo is $t \sim
96 P_{44}^{-1}$ Myr, if $\Gamma = 5/3$;  this increases to  $150 /
P_{44}$ Myr if $\Gamma \ 4/3$.  We can  estimate the current expansion
rate of  the outer edge, from equation (3).  This gives  $ d R / dt
\sim 100 P_{44}$ km/s  if  $\Gamma = 4/3$, and  $\sim 160$   
km/s if $\Gamma = 5/3$.  For comparison, the sound speed in the
external gas is $c_s \sim 400$ km/s, so that the expansion is
subsonic.  The internal energy of the bubble at time $t$ is $U_{int} =
( 2.2 / 3 \Gamma) P_j t$.  Thus, a fraction $0.44$ of the total jet
power goes to internal energy if $\Gamma = 5/3$;  this fraction
becomes $0.55$ if $\Gamma = 4/3$.  The rest of the jet power goes to
$p dV$ work (and a small fraction to radiation).

\subsection{Further Issues}

To summarize, we find that the radio halo is still alive, rather than
being a relic of previous activity.  It is 
 expanding at about 1/4 of the local sound speed, and is
significantly younger than the galaxy that hosts it.  However, this is 
not the end of the story;  several interesting questions remain
unanswered.

What is the origin of the bright radio filaments?  There are at
least three possibilities.  The first is that they are highly
compressed regions, or shocks, in transonic turbulence.  The second
is that they are the intermittent magnetic flux tubes seen in MHD
turbulence at even low turbulent Mach numbers \citep{ki95}.  The third is
that they are similar to the filaments seen in numerical simulations
of passive-field radio sources \citep{cl96}.  All three possibilities
will lead to regions of high field and possibly high particle density,
which will be bright in synchrotron radiation. 

What is the nature of the ``ear'' at the east edge of the
source?  DeYoung (1998, private communication) pointed out to us that
it resembles a subsonic vortex ring.  This is an interesting
possibility.  Little is known about vortex rings in compressible,
magnetized plasmas, but one might expect them to retain their basic
characteristics.  Its reason for existence, however, remains unclear.
Further, are  the east and west flows fundamentally different, so that
one disrupts and one creates a vortex ring, or is the different
environmental? 

Is {\it in situ} electron acceleration necessary?  In the bright
filaments, with  $B \sim 10 \mu$G, electrons radiating at 1 GHz live
only $\sim 50$ Myr, which is a bit less than the age of the halo.
This result suggests either that the electrons undergo {\it in situ}
re-acceleration in the filaments, or else that they spend most of
their life in a weaker magnetic field, possibly in the interfilament
region.  

 What is the nature of the outer edge, and how efficient is the
mixing across it, of external gas with radio-loud plasma? The X-ray
profile, together with the apparent integrity of the radio edge,
suggests there has been partial mixing.  Should we
look to the earth's magnetopause for an analogy (where patchy reconnection
allows some fraction of the incoming plasma to mix across the
boundary)?  This may be a good analogy if the edge of the bubble is
defined by a quasi-parallel magnetic field. Alternatively, should we
look to the outer edges of  old supernova remnants (which have highly
filamented outer shells, and are thus more ``open'' to mixing with the
ambient gas)? This may be a better analogy of the edge of the bubble
is governed by fluid, not magnetic, processes.  

 Finally, why is this radio galaxy so unusual? Most radio
galaxies at this power are supported by jets which remain collimated
and directed for scales $\sim 10 - 100$ kpc.  However, the M87 jet
continues undisturbed for only a few kpc.   What is the
reason for the jet disruption so close the galactic
center?  Is it connected somehow to the dense cooling core in which
the source sits, and with the existence of a small set of amorphous
central radio galaxies in other, similar, cooling-core clusters?

\section{Physical Picture:  The Cluster Core}

The Virgo X-ray halo is smooth, centrally peaked,  and nearly circular
(ellipticity $\sim 0.1$ on large scales \citep{fg83}. Using spherical 
deprojection of ROSAT data \citep[also Table 2]{nb95}, one
 derives densities and  temperatures
which suggest that the cooling time is less  than the
Hubble time  for the inner $\sim 80$ kpc.  
The smooth X-ray structure and short central cooling time have  
led various authors to suggest that M87 sits in the center of a modest
cooling flow.   In the simple cooling-flow model, radiative losses 
are balanced by slow settling of the  gas in the cluster
gravitational potential ({\it e.g.}, \citep{f94}).     
For M87, the cooling inflow has been estimated as 
 $\dot M \sim 24 M_{\sun}$/yr ({\it e.g.} \citet {p98}, again scaled
to 17 Mpc). Such a smooth,  spherically symmetric inflow, with the
densities given in Table 2,  will be very slow:  it will have inflow
velocity $\sim 2$ km/s at 50 kpc. 

\subsection{Activity in the Core}

Consideration of the radio data shows that the picture is more complex 
than a simple cooling-driven inflow. 
This is demonstrated by several pieces of evidence. 

{\it The Radio Image} itself shows evidence of a disturbed inner halo.  The
filaments suggest both directed and disordered flows.  These flows are
magnetized and possibly transonic (which we suggest based on the enhanced
radio emission from the filaments, which are probably regions of 
strong magnetic field).

{\it High-Resolution X-Ray Data} \citep{f87,nb95} show 
complex structure in the inner $\sim 20$ kpc of the X-ray emitting
gas.  This gas  has a very asymmetric distribution, with approximately
the same  extent as the radio
halo.  	In addition, new work by \citep{h99} shows that the X-ray excess
is enhanced in a narrow ridge which crosses the jet close to knot A,
the brightest knot in optical and radio.  This strongly
suggests that the jet and the X-ray gas are interacting on this scale. 

{\it The Dynamic Radio Halo} is being supplied with energy from the jet.  
It should be expanding outwards, into the thermal gas, driven by its own 
internal energy.  Our simple model
predicts an expansion speed $\sim 100$ km/s, which is 
approximately 1/4 of the local sound speed.  This must have an impact 
on the ambient cluster gas.

{\it The Inner Few kpc} of M87, at least, are magnetized and turbulent.  
Emission line clouds embedded in the gas are moving at
$\sim 200$ km/s  \citep{s93,k96}.  It is very likely that these clouds
share the velocity field of the hotter gas in which they sit.  	In
addition, Faraday rotation data reveal dynamically significant,
ordered magnetic fields in this region \citep{o90,z98}.  The magnetic
field is probably supported by transonic
turbulence \citep{e99}.  Taken together, these observations suggests that
the very inner part of the M87 halo is  
turbulent, probably at transonic levels.  

\subsection{Energetics and Activity}

The evidence shows that the Virgo core is a complex place. 
 On scales $\ltw 40$ kpc, it
contains a dynamic mixture of relativistic particles, thermal plasma
and magnetic fields.  We suspect that the entire halo is similar to the
inner few kpc, being turbulent and magnetized at significant levels.

It follows that this region cannot be described as a simple cooling
flow.  This can be demonstrated by basic energetics.
The bolometric X-ray luminosity from the region occupied by
the radio source,  roughly the central 40 kpc, is $L_x \sim 0.9 \times
10^{43}$erg/s. (To derive this, we use the densities and temperatures
from Table 2, with the cooling curve given by Westbury and Henriksen
1992).  The 
central cooling can also be found using the value given by \citet{p98}
for $L_x$ within cooling core, scaling that value to 17 Mpc, and then
using the deprojected density  structure from Table 2 to find the
luminosity within 40 kpc).  In the absence of a comparable energy
input, the central gas must indeed cool, and collapse in the
gravitational well of the system.  

However, the picture is not so simple for Virgo.  The black hole in
the galactic nucleus is currently supplying $P_j > 10^{44}$ 
erg/s  to the radio halo.
Only a small fraction of this is leaked to radio
emission. The bulk of this energy must be deposited in kinetic or
internal energy of the composite plasma in the Virgo core.  This input is
at least as large, and probably much larger, than the radiative losses
from the core.  It follows that the region is more complex than a simple
cooling flow.  \cite{bi99} has already pointed this out, and  
we agree with him.  

What effects might we expect? 
The energy from the core is being deposited on a scale
$\sim 40$ kpc, throughout the radio halo.  Some fraction of the jet
energy will go to turbulence or bulk flows (such as the 
$\sim 100$ km/s expansion of the bubble), while the rest will  
go directly to heating.  The heat will directly drive an outwards
expansion of the heated gas into the cooler, lower-pressure gas
sitting above it.  The turbulence will enhance magnetic fields,
through dynamo effects, which may in turn enhance the turbulence
through magnetic buoyancy.  We cannot predict the exact fraction of
energy deposited in flows or in heat; recalling the energetics
estimated in \S 3.2.2, we might estimate that approximately half the
energy goes into each mode. 

The thermal state of the gas will be determined by all of these
competing processes --- heating, radiative cooling, thermal
conduction, expansion, turbulence
and turbulent dissipation.  X-ray data find a component of gas at $\sim 1.1$
keV in the inner arcmin, and warmer gas at $\sim 3$ keV outside of $\sim 3$
arcmin \citep{fg83, nb95, ma99}. Our simple arguments here cannot predict the
full temperature structure of the core.  Previous steady-state, 
spherically symmetric models \citep{tt81, tr83}, provided interesting
first attempts but also cannot address the complexity of the region.
More detailed, time-dependent simulations will be required.  

In the absence of full numerical  simulations, simple lifetime
considerations can be  helpful.   The current energy content within
$\lesssim 40$ kpc is  $U_{th} \sim 3 \times 10^{59}$ erg (again using
data from Table 2).  If the jet maintains  an average energy input of
$10^{44} P_{44}$ ergs s$^{-1}$, we can expect something like half of
that power to be deposited in the cluster gas, perhaps shared equally
between direct heating, and turbulence (which will eventually decay
into indirect heating). It follows that it will take $\sim 190 /
P_{44}$ Myr 
to deposit an amount of energy comparable to that currently in the region.
For comparison, the radiative lifetime of the region is $U_{th} / L_x
\sim 1$ Gyr.  Thus, based on our estimated lower limit for the jet
power, its energy input clearly dominates radiative losses at
present.  We estimated above that the current age of the radio halo
is $\sim 100-150 / P_{44}$ Myr.   Thus, the radio halo is young
compared to the age of the galaxy,  but ``just right'' compared to the
energy turnover times in the Virgo core.   

This suggests to us an on-off duty cycle for
the central engine in M87.  The engine may have been active for only
100-200 Myr, which happens to be comparable to the energy turnover
times for the core.   As long as the engine remains active at its current
level, it will have a strong effect on the local cluster core.  It
will heat the core and support turbulence, bulk flows and magnetic
fields in that core.  These extra pressure sources will in turn offset
radiative cooling, support the gas against gravitational infall, and
possibly drive the central regions outwards in a local expansion. Once
the central engine turns off, the gas will cool and return to
a quasi-static state as the turbulence and bulk flows dissipate.  \cite{bt95}
have simulated a similar situation for a smaller galaxy (with the 
differences that they assumed the energy was all deposited at the very
center of the flow, and in their spherically symmetric model they
could not include turbulent flows or magnetic fields).  Their calculation
shows the cyclic response of the ambient gas as it heats and cools,
expands and collapses, in response to the activity cycle of the central
energy source.

\section{Conclusions}

We have obtained new and wonderful radio images of Virgo A,  the
large-scale M87 radio halo.  Our images clearly show that the halo is
a complex and active object.  Comparison of our data to recent
X-ray images shows that the radio-loud and X-ray-loud plasmas are 
interacting strongly in the core of the Virgo cluster.  Simple
energetics shows that the input to the region, from the active
nucleus in M87, exceeds radiative losses at the present epoch. 

We thus arrive at a new picture of M87 and the Virgo Cluster.
We find a competition for dominance between
the slow, cooling-driven inflow of the hot cluster gas,
and the violent outflow of
energy from the black hole in the galactic nucleus.  At present the
black hole appears to be winning in the inner region of the system.
The detailed processes taking place on scales from a few kpc to
40 kpc are likely to arise from  this complex competition. 
However, both the timescales
we have derived and the general wisdom about radio galaxy lifetimes 
suggest that the current outflow is likely to be  transient.
If the AGN activity diminishes or stops altogether, then the physical
processes which can create a cooling flow would dominate.  It may be that M87, 
like other central active galaxies in cooling cores, has gone 
through several heating/cooling cycles during its lifetime.  

\begin{acknowledgements}
We thank our referee, Dave De Young, for insightful comments.  
JE has been partially supported by NSF grant AST-9720263.  Basic
research at the Naval Research Laboratory is supported by the Office
of Naval Research. 
\end{acknowledgements}

\clearpage

\figcaption[m871.eps]{90cm Image of M87. The central region containing the
jet and inner radio-lobes is in the red-orange region near the image center.
The convolving beam for this Maximum Entropy Image
is $7.8\times6.2$ pa=86 degrees.  One arcsec corresponds to 85 pc
at our assumed distance of 17 Mpc.
\label{m87color}}

\figcaption[m872.eps]{90cm Image of central part of Virgo A/M87 
radio halo shown in figure 1. This
display shows the sharp surface brightness drop between the
inner radio-lobe structure (which contains the galactic nucleus and the
well known jet; solid black in this figure) and the east and west outflows
which emerge from the inner lobes.
\label{m87pc}}

\figcaption[m873.eps]{Greyscale Image of Virgo A/M87  and the
Location of Minimum Pressure Measurements given in table 1.
\label{minpfig}}
\clearpage

\onecolumn
\begin{table}[htb]
\caption{Minimum Pressure Analysis for M87 Halo features\label{minp}}
\vspace{8pt}
\begin{minipage}{6.2in}
\begin{center}
\begin{tabular}{cccccccc}
\noalign{\medskip}
\strut Feature  & $\Theta_{proj}$\tablenotemark{1} & $D_{proj}$\tablenotemark{2}
 & $f(\nu_o)$\tablenotemark{3} & $W$\tablenotemark{4}
        & $B_{min~ p}$\tablenotemark{5} & $p_{min~ p}\tablenotemark{6} $
\\
\strut   & (asec)& (kpc) & mJy/beam & (asec)
        & ($\mu$G) &  ( $10^{-11}$dyn/cm$^2$)
\\
\noalign{\medskip}
\noalign{\medskip}
A & 210 & 18 & 25.0 & 24.5 & 7.48 & .52
\\
B & 203 & 17 & 52.9 & 15.4 & 10.6 & 1.0
\\
C & 68 & 5.8 & 58.8 & 41.4 & 8.22 & .62
\\
D & 115 & 9.8 & 70.6 & 33.3 & 9.19 & .78
\\
E & 157 & 13 & 13.8 & 8.8 & 8.46 & .66
\\
F & 177 & 15 & 24 & 14 & 8.70 & .70
\\
G & 255 & 30 & 28.4 & 19 & 8.38 & .65
\\
H & 334 & 28 & 20 & 26 & 6.90 & .44
\\
 
\end{tabular}
\end{center}
\end{minipage}
\tablenotetext{1}{$\Theta_{proj}$ is the angular projected distance of the
feature from the core.}
\tablenotetext{2}{$D_{proj}$ is the linear projected distance of the
feature from the core.}
\tablenotetext{3}{$f(\nu_o)$ is the surface brightness of the feature}
\tablenotetext{4}{$W$ is the width (approximately the full width at zero power)
 of the feature.}
\tablenotetext{5}{$B_{min~ p}$ is the field from the minimum pressure
solution}
\tablenotetext{6}{$p_{min~p}$ is the pressure from the minimum pressure
solution}

\end{table}
\clearpage

\begin{table}[htb]
\caption{X-ray density and pressure profile \label{xrayp}}
\vspace{8pt}
\begin{minipage}{6.2in}
\begin{center}
\begin{tabular}{cccccccc}
\noalign{\medskip}
\strut   & Scale & distance & $n_e$    & $k _BT$ & $p$
\\
\strut   & (amin)& (kpc) & $10^{-3}$cm$^{-3}$ & keV
        & $10^{-11}$dyn/cm$^2$
\\
\noalign{\medskip}
\noalign{\medskip}
&0.83  & 4.1    & 82.4  & 1.10  & 14.5
\\
&1.67  & 8.2    & 32.6  & 1.14  & 5.95
\\
&2.50  & 12.3   & 18.7  & 1.41  & 4.27
\\
&3.33  & 16.4   & 14.7  & 1.36  & 3.20
\\
&4.17  & 20.6   & 11.3  & 2.3  & 4.17
\\
&5.0  & 24.7    & 9.06  & 2.0  & 2.89
\\
&5.83  & 28.8   & 7.53  & 1.7  & 1.99
\\
&6.67  & 32.9   & 6.56  & 1.6  & 1.67
\\
 
\end{tabular}
\end{center}
\end{minipage}
\tablenotetext{}{Data from Nulsen \& Boheringer (1995), scaled to our
assumed distance of 17 Mpc.} 
\end{table}
\clearpage
\end{document}